\documentclass[sigconf]{acmart}
\author{Zhenxiang Xu}
\authornotemark[2]
\authornotemark[3]
\affiliation{
\institution{Zhejiang University}
\city{Hangzhou}
\country{China}
}
\email{zhenxiangxu@zju.edu.cn}

\author{Jiawei Chen}
\authornote{Corresponding author.}
\authornote{State Key Laboratory of Blockchain and Data Security, Zhejiang University.}
\authornote{College of Computer Science and Technology, Zhejiang University.}
\authornote{Hangzhou High-Tech Zone (Binjiang) Institute of Blockchain and Data Security.}
\affiliation{
\institution{Zhejiang University}
\city{Hangzhou}
\country{China}
}
\email{sleepyhunt@zju.edu.cn}

\author{Sirui Chen}
\authornotemark[2]
\authornotemark[3]
\affiliation{
\institution{Zhejiang University}
\city{Hangzhou}
\country{China}
}
\email{chenthree@zju.edu.cn}

\author{Yong He}
\affiliation{
\institution{Ant Group}
\city{Hangzhou}
\country{China}
}
\email{heyong.h@antgroup.com}

\author{Jieyu Yang}
\affiliation{
\institution{Ant Group}
\city{Hangzhou}
\country{China}
}
\email{jieyu.yjy@antgroup.com}

\author{Chuan Yuan}
\affiliation{
\institution{Ant Group}
\city{Hangzhou}
\country{China}
}
\email{yuanzheng.xy@antgroup.com}

\author{Ke Ding}
\affiliation{
\institution{Ant Group}
\city{Hangzhou}
\country{China}
}
\email{dingke.dk@antgroup.com}

\author{Can Wang}
\authornotemark[2]
\authornotemark[4]
\affiliation{
\institution{Zhejiang University}
\city{Hangzhou}
\country{China}
}
\email{wcan@zju.edu.cn}
\AtBeginDocument{%
  }

\setcopyright{acmlicensed}
\copyrightyear{2018}
\acmYear{2018}
\acmDOI{XXXXXXX.XXXXXXX}
\acmConference[Conference acronym 'XX]{Make sure to enter the correct
  conference title from your rights confirmation email}{June 03--05,
  2018}{Woodstock, NY}
\acmISBN{978-1-4503-XXXX-X/2018/06}

\newcommand{\ie}{\emph{i.e., }}
\newcommand{\eg}{\emph{e.g., }}

\newcommand{\cf}{\emph{cf. }}

\usepackage{amsthm, amsmath, amssymb}
\usepackage{pifont}
\usepackage{tabularx}
\usepackage{enumitem}
\usepackage{booktabs}
\usepackage{multirow} 
\usepackage{graphicx}
\usepackage{amssymb} 
\usepackage{xspace}
\usepackage{xcolor}
\usepackage[title]{appendix}
\newcommand{\ours}{TrieRec\xspace}
\begin{document}

\title{Trie-Aware Transformers for Generative Recommendation}



\begin{abstract}
Generative recommendation (GR) aligns with advances in generative AI by casting next‑item prediction as token‑level generation rather than score-based ranking. Most GR methods adopt a two‑stage pipeline: (i) \textit{item tokenization}, which maps each item to a sequence of discrete, hierarchically organized tokens; and (ii) \textit{autoregressive generation}, which predicts the next item's tokens conditioned on the tokens of user’s interaction history. Although hierarchical tokenization induces a prefix tree (trie) over items, standard autoregressive modeling with conventional Transformers often flattens item tokens into a linear stream and overlooks the underlying topology.

To address this, we propose TrieRec, a trie‑aware generative recommendation method that augments Transformers with structural inductive biases via two positional encodings. First, a \textit{trie‑aware absolute positional encoding}  aggregates a token's (node's) local structural context (\eg depth, ancestors, and descendants) into the token representation. Second, a \textit{topology‑aware relative positional encoding} injects pairwise structural relations into self‑attention to capture topology‑induced semantic relatedness.  TrieRec is also model‑agnostic, efficient, and hyperparameter‑free. In our experiments, we implement TrieRec within three representative GR backbones, achieving notably improvements of 8.83\% on average across four real-world datasets.

\end{abstract}




\received{20 February 2007}
\received[revised]{12 March 2009}
\received[accepted]{5 June 2009}

\maketitle

\section{Introduction}

\begin{figure}[t]
    \centering
    \includegraphics[width=\linewidth]{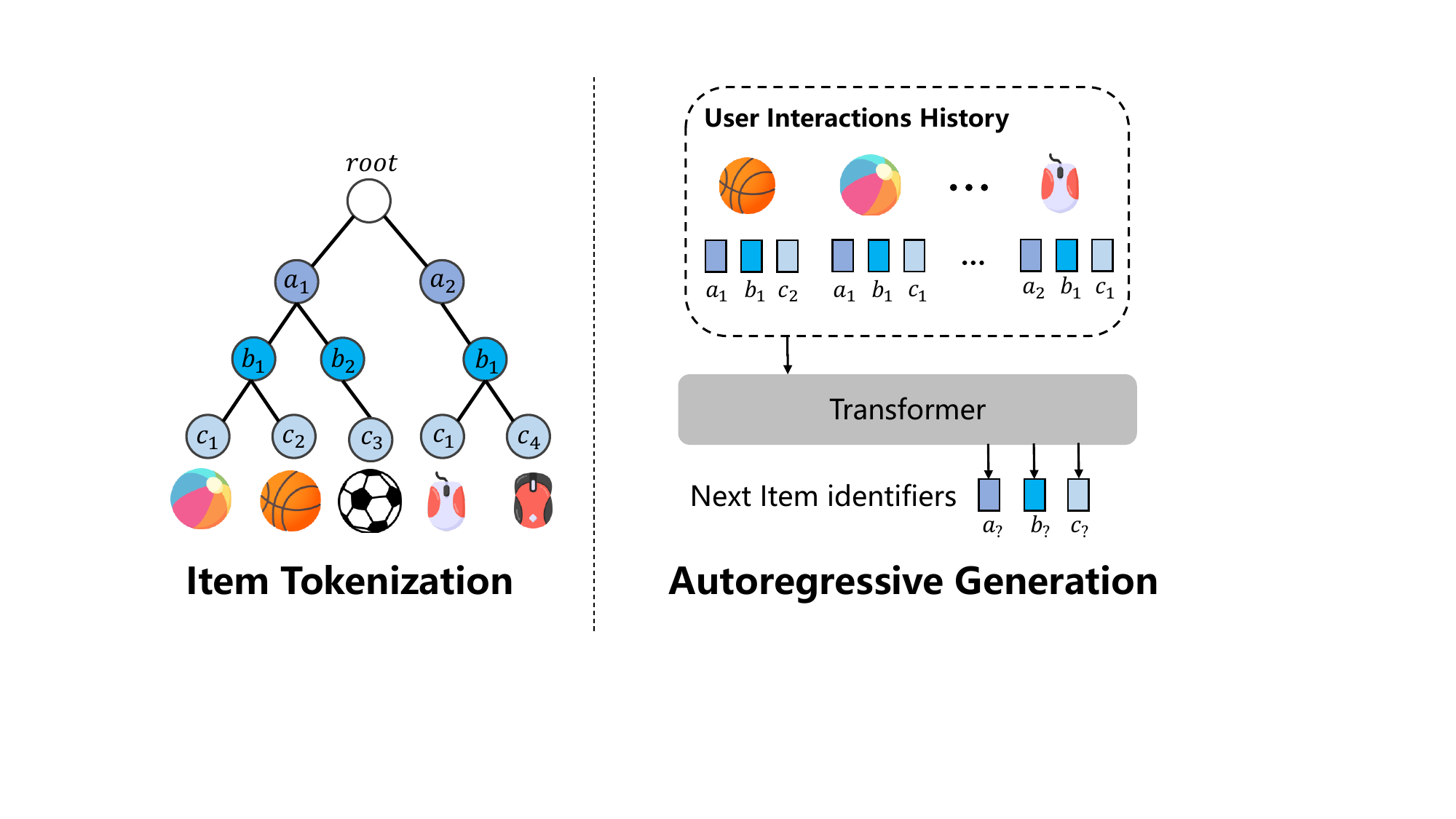}
    \caption{Generative recommendation pipeline. Item tokenization maps items to semantic token sequences via hierarchical quantization, forming a trie structure. Autoregressive generation employs a Transformer to predict the next item's token sequence conditioned on user interaction history.}
    \label{fig:intro}
\end{figure}

Motivated by recent advances in generative AI \cite{gpt3,gpt4}, \textit{Generative Recommendation} (GR) has recently emerged as a promising paradigm for recommender systems \cite{zhao2024recommender,deldjoo2024review}. Unlike discriminative approaches that score query–item pairs \cite{koren2009matrix,he2017neural,sasrec},  GR directly generates target items in an end-to-end generative manner \cite{tiger,letter}. A typical GR pipeline comprises two stages \cite{li2025survey,wang2023generative}:  (i) item tokenization, which learns a codebook that maps each item to a sequence of semantic identifiers (discrete tokens); and (ii) autoregressive generation, which predicts the next token conditioned on a token sequence constructed from the user’s interaction history. In practice, item tokenization is often realized with hierarchical quantization or clustering (\eg RQ-VAE\cite{rqvae}, RQ-kmeans \cite{luo2025qarm}), producing coarse-to-fine semantic tokens. These tokens naturally form a \textit{hierarchical prefix tree }\cite{vijayakumar2016diverse,zhuo2020learning} (trie, \cf Figue \ref{fig:intro}), where each item corresponds to a unique path from root to leaf. Consequently, GR can be viewed as layer-wise child-node (next-token) selection along the trie, rather than direct retrieval from the entire item corpus, thus restricting the per-step search space and keeping computation affordable \cite{bao2024decoding,wang2025msl}. This mechanism parallels the principles of conventional cascade recommendation systems \cite{covington2016deep}, which progressively narrow candidates from coarse to fine via multiple models; while GR advances this concept by unifying the process within a single end-to-end generative framework. 

Despite recent progress, existing GR approaches suffer from a critical limitation: \textit{the inherent tree structure of item tokens is often overlooked during autoregressive modeling.} Specifically, existing GR models mainly employ Transformers \cite{attention} that flattens hierarchical token sequences into a linear stream, relying solely on self-attention mechanisms to learn token semantics and dependencies. This neglects the rich topological and relational information encoded in the trie structure. In fact, the trie fundamentally provides two sources of valuable knowledge: (i) \textit{Structural position}, which situates a token by its depth and item path and anchors it in the surrounding hierarchical context. Considering a token (identifier) may be reused across items, cues from ancestors and descendants disambiguate its semantics and enhance token modeling. (ii) \textit{Relational topology}, which captures structure-aware relations among tokens (\eg ancestor–descendant links and shared-prefix siblings), inducing different degrees and types of semantic relatedness. For example, tokens along the same root-to-leaf path belong to a single item and typically exhibit strong semantic dependency;  whereas tokens that share a closer lowest common ancestor (LCA) \cite{bender2000lca} tend to originate from similar items and are more closely related than tokens from distant branches of the trie.

To bridge this gap, we propose TrieRec, a trie-aware generative recommendation method that augments Transformers with structural inductive bias via tailored positional encodings:

\begin{itemize}[leftmargin=*]

\item \textbf{Trie-aware absolute positional encoding}: a hierarchy-dependent neural encoder that aggregates a node's local structural context (\eg depth, ancestors and descendants) into an embedding added to the token representation. This equips tokens with awareness of their hierarchical position and item membership, and directly incorporates useful information from structurally related nodes.

\item \textbf{Topology-aware relative positional encoding}: a pairwise structural bias for attention that encode topological relations between token pairs. Specifically, for each node pair, we compute their lowest common ancestor (LCA) and the distances from each token to the LCA, and map these features to a learnable attention-bias term to capture topological-induced semantic relatedness.

\end{itemize}

Beyond structural awareness, TrieRec also offers practical advantages: (i) model‑agnostic: easily plugged into a variety of GR backbones; 
(ii) efficiency: limited additional computational overhead;  
(iii) hyperparameter‑free: no new tunable hyperparameters. To empirically validate TrieRec, we integrate it into three representative GR backbones (including TIGER, CoST, LETTER), achieving an average relative improvement of 8.83\% on four real‑world datasets.

In summary, the contributions of this work are:

\begin{itemize}[leftmargin=*]

\item We highlight the critical importance of explicitly encoding the trie structure in the autoregressive modeling of generative recommendation.

\item We propose TrieRec, which augments Transformers with two tailored positional encodings, capturing both tokens' structural positions and pairwise topological relations in the trie. 

\item We conduct extensive experiments to validate the effectiveness of TrieRec, showing notable improvements over state-of-the-art baselines with limited additional computational cost. 

\end{itemize}

\section{Preliminary}\label{sec:preliminary}

\textbf{Problem Definition.}
Following recent work~\cite{tiger,letter,cost}, we also focuses on {sequential recommendation}~\cite{sasrec,bert4rec}, a widely adopted setting in real-world recommender systems. Let $\mathcal{I}$ denote the item set. Given a user’s historical interaction sequence  $S = \{ i_{1}, i_{2}, \dots, i_{T} \}$, where $i_t \in \mathcal I$ denotes the $t$-th item in the sequence, the goal is to predict the next item $i_{T+1}$ that the user is most likely to interact with.

\textbf{Generative Recommendation.}
Recent advances in generative AI have stimulated extensive interest in \emph{generative recommendation} (GR).
Different from discriminative recommenders that estimate matching scores for query-item pairs, GR formulates recommendation as an autoregressive generation problem, where the model directly generates the next item in a tokenized space.
Most GR methods follow a two-stage pipeline \cite{tiger,letter,cost}:

\textit{Item Tokenization.}
This stage encodes each item $i$ into a sequence of structured discrete tokens (\ie semantic identifiers) $\mathbf{c}_i = [c_{i,1}, c_{i,2}, \dots, c_{i,L}]$.
A representative approach is RQ-VAE \cite{rqvae}, which recursively quantizes item embeddings from coarse to fine granularity.
Recent studies have further explored improved tokenization strategies, such as collaborative-enhanced or learnable tokenization \cite{letter}, while often adhering to the hierarchical encoding principle.

Such hierarchical tokens naturally induce a \emph{prefix tree} (trie; cf.~Figure~\ref{fig:intro}), where each item corresponds to a unique root-to-leaf path and different items may share common prefix tokens. Such hierarchical trie-structured representation offers several advantages: 1) The tree structure is well suited for generative modeling. Intuitively, generation can be interpreted as {layer-wise child-node (next-token) selection} along the trie, which restricts the search space at each step and keeps modeling computationally manageable. 2) The tree structure facilitates collaborative semantic modeling. items sharing longer prefix paths tend to be semantically closer, and the trie explicitly captures such relatedness in the shared prefix token generative probability.

\textit{Autoregressive Generation.}
This stage models the token sequence of the next item in an autoregressive manner.  Specifically, a user’s historical interactions are represented as a token stream:
\[
\mathbf{X} = [c_{1,1}, c_{1,2}, \dots, c_{1,L}, \dots, c_{T,1}, c_{T,2}, \dots, c_{T,L}],
\]
where $(c_{t,1}, c_{t,2}, \dots, c_{t,L})$ denotes the token sequence of the $t$-th interacted item.
A GR model $\mathcal{M}$ is then employed to model the probability of the next-item token sequence conditioned on the history:
\begin{equation}
P_{\mathcal{M}}(\mathbf{y} \mid \mathbf{X})= \prod_{t=1}^{L} P_{\mathcal{M}}(y_t \mid y_{<t}, \mathbf{X}).
\label{eq:gr_factorization}
\end{equation}
where $\mathbf y=[y_1,y_2,...,y_L]$ denotes the token sequence of the predicted next item. The generative probability of the next item  factorizes into the product of token-level conditionals. 

Given the empirical successes of Transformer architectures in sequence modeling, Transformer-based backbones are commonly adopted in GR. The Transformer relies on self-attention to model token dependencies, which can be summarized as follows:
\begin{align}
    \mathbf{o}_u &= \sum_{v} A_{uv} (\mathbf{e}_v \mathbf{W}^V) \\
    A_{uv} &= \text{Softmax}_v \left( \frac{(\mathbf{e}_u \mathbf{W}^Q)(\mathbf{e}_v \mathbf{W}^K)^\top}{\sqrt{d}} \right) 
\end{align}
where $\mathbf{e}_u, \mathbf{e}_v \in \mathbb{R}^d$ are the embedding of tokens at positions $u$ and $v$ respectively, and $W^Q$,$W^K$,$W^V$ are learnable projection matrices. Each token $u$ aggregates information from related context tokens $v$ in the sequence according to the attention weights $A_{uv}$, which reflect semantic dependency measured by the similarity of the transformed token embeddings, resulting in the aggregated output representation $\mathbf{o}_u$.. While vanilla Transformers are powerful, they are agnostic to token order. Linear positional encodings  are typically employed. For example, the conventional  model used by TIGER encodes the token distance within the sequence \cite{t5}.

\textbf{Limitations of Existing GR Methods.}
Although the item tokens are often organized as a trie, existing GR models often treat the token stream $\mathbf{X}$ as a flat sequence during autoregressive generation, without explicitly modeling the underlying trie structure. This omission discards two types of crucial structural signals:

\emph{(i) Absolute structural position in trie.}
A token's position in the trie, signifies its depth and root-to-leaf path, as well as anchors it within the hierarchical context.
This information is particularly important because token identifiers can be reused across different items. For instance, under RQ-VAE-style tokenization, the same token may just reflect quantized residuals rather than a specific fixed semantic meaning as in natural language vocabularies. Therefore, incorporating cues from ancestor and descendant nodes can help disambiguate token semantics and improve token semantic modeling.

\emph{(ii) Relational topology between token pairs.} The structural relations in the trie encode varying strengths and types of semantic relatedness. For example, tokens along the same root-to-leaf path belong to the same item and typically exhibit strong semantic dependency. Their hierarchical relationships also exhibit different patterns. Moreover, token pairs with a closer lowest common ancestor (LCA) (\eg token $c_2$ of item \textit{basketball} and token $c_3$ of item \textit{football}) tend to originate from more similar items and are more related than those located in distant branches of the trie (\eg tokens $c_2$ of item \textit{basketball} and token $c_1$ of item \textit{mouse}).

Although the trie structure is central to tokenization, its potential is underexplored in autoregressive generation. To the best of our knowledge, only a single work has attempted to incorporate trie information into a contrastive loss objective. However, their approach is not well aligned with contemporary generative modeling, as it requires computing and storing embeddings for every tree node, incurring substantial time and memory costs. Moreover, their modeling only focuses on local parent–child relationships rather than exploiting token positions and higher-order relations. Our empirical investigations also reveal their performance is suboptimal (\cf section \ref{sec:overall}).

\section{Methodology}
In this section, we present TrieRec, a trie-aware generative recommendation framework. TrieRec introduces two positional encodings:  trie‑aware absolute positional encoding (Subsection \ref{sec:tae}) and topology‑aware relative positional encoding (Subsection \ref{sec:tre}). We then discuss practical advantages of TrieRec and relate it to relevant prior work (Subsection \ref{sec:discussion}).
\begin{figure*}[htbp]
    \centering
    \includegraphics[width=\linewidth]{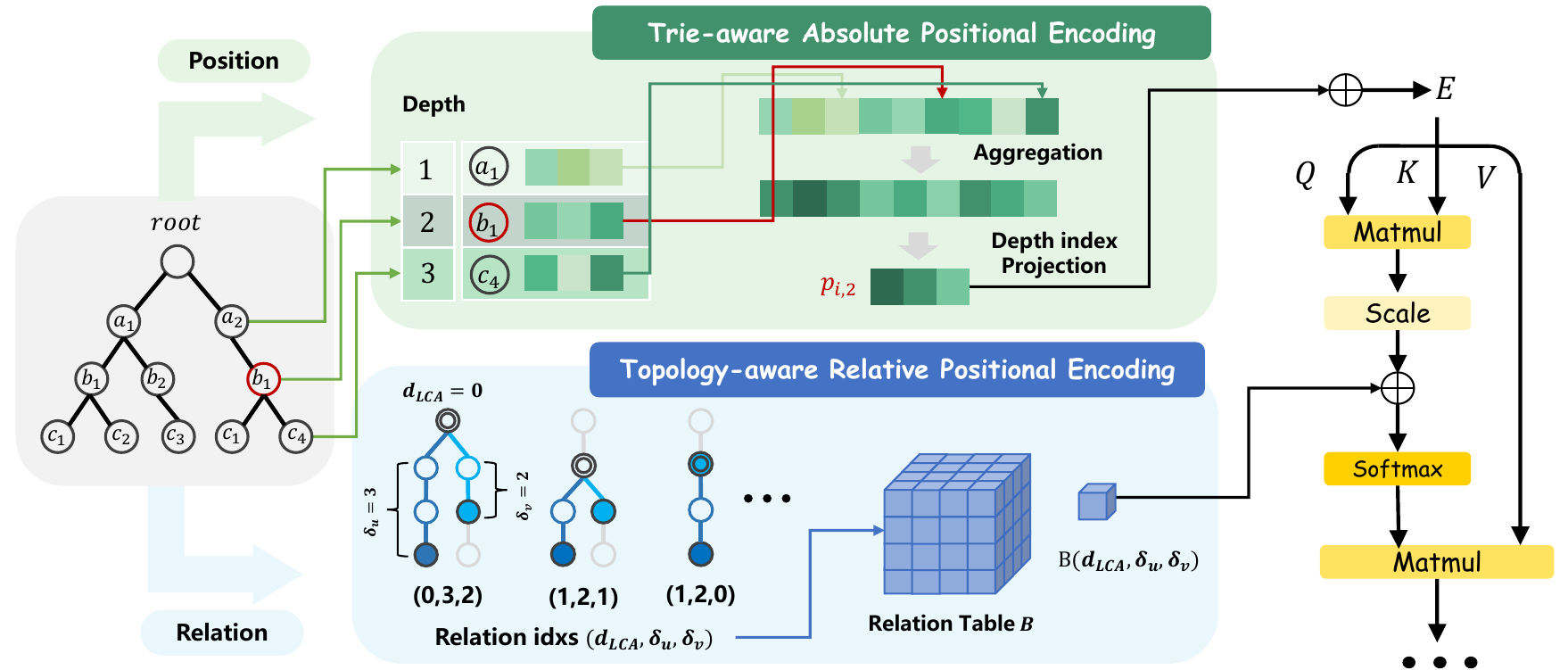}
    \caption{The illustration of our proposed trie-aware absolute positional encoding and topology-aware relative positional encoding in \ours.}
    \label{fig:method}
\end{figure*}
\subsection{Trie‑aware Absolute Positional Encoding (TAE)}
\label{sec:tae}
 Token structural positional information provides rich semantic signals, indicating where a token resides in the hierarchical tree (its layer and path) and anchoring it within the structural context.  To enrich token embeddings with such structural information, we define a dedicated neural encoder that generates a position encoding for each token $c_{i,l}$ as follows:
\begin{align}
    \mathbf{x}_i^{con} &\overset{\text{\ding{172}}}{=} [\mathbf{e}_{i,1}^\top; \mathbf{e}_{i,2}^\top; \dots; \mathbf{e}_{i,L}^\top]^\top \in \mathbb{R}^{L \cdot d} \\
    \hat{\mathbf{x}}_i &\overset{\text{\ding{173}}}{=} \text{MLP}(\mathbf{x}_i^{con}) \in \mathbb{R}^{L \cdot d} \\
    \mathbf{p}_{i,l} &\overset{\text{\ding{174}}}{=} \mathbf{W}_{l} \hat{\mathbf{x}}_i + \mathbf{b}_{l} \quad l=1, \dots, L
\end{align}
where $e_{i,l}$ denotes the original semantic embedding of token $c_{i,l}$ at a given Transformer layer. Our encoder consists of three stages: \ding{172} Concatenation: We first collect the embeddings of the token’s hierarchical context neighbors, including ancestors and descendants ($c_{i,1}, c_{i,2}, ..., c_{i,L}$), and concatenate them into a flat vector $x_i^{con}$. \ding{173} Aggregation: We pass $x_i^{con}$ through a MLP to model information aggregation and interaction among its components. \ding{174} Projection: We apply a hierachy-aware projector to generate token positional encodings based on the token layer index.

The rationale of these designs can be explained as follows: 1)  In a tree, a node’s (token’s) position is uniquely determined by the sequence of its ancestors. Accordingly, we explicitly encode the full longitudinal context to capture positional information. Descendant information belonging to the same item is also included. Intuitively, those tokens belonging to same item path can naturally share closely semantic dependency, offering item‑level semantics and benefiting token semantic modeling. 2) Unlike permutation-invariant attention, an MLP is position-dependent: each dimension of $x^{path}_{i,l}$ has its own learnable weights to aggregate information across dimensions, leveraging the hierarchical positions of tokens. This also enables simultaneous cross-token and cross-dimension interactions, allowing more fine-grained information aggregation. 3) To further capture different semantic properties across token layers, a hierarchy-aware projector generates the token positional encoding using different parameters for different layers, thereby modeling their distinct characteristics.

The learned position encoding $\mathbf{p}_{i,l}$ is further added to the original token embedding $\mathbf{e}_{i,l}$ in the transformer, enriching its representation with the structural context:

\begin{equation}
\hat{\mathbf{e}}_{i,l}=\mathbf{e}_{i,l} + \mathbf{p}_{i,l}
\label{eq:tae}
\end{equation}

In practice,  The MLP can be implemented as a simple two-layer network with an activation function $\text{GELU}$ and layer normalization. For efficiency, the TAE is applied only in the first transformer layer, and it is not added for generated tokens, since the complete item-path information for these tokens is not available. We find that this lightweight implementation has yielded meaningful improvements with minimal overhead.

\subsection{Topology-aware Relative Positional Encoding (TRE)}
\label{sec:tre}
We further exploit the  topological patterns of token pairs. Different topological relations convey different strengths and kinds of semantic dependencies.  We adopt two guiding principles: (i) Topology-aware design: explicitly model multiple topological features (\eg structural distance, the lowest common ancestor, and relative hierarchical relations) for token pairs. (ii) Computational efficiency: keep the design concise and scalable, avoiding heavy per-pair computations.

To achieve this, for every pair of tokens $(u, v)$ within a sequence, we compute three topological features: (i) $d^{LCA}_{u,v}$: the tree depth of their LCA ; (ii) $\delta_u$: the distances from $u$ to the LCA; (iii) $\delta_v:$ the distance from $v$ to the LCA. These three features jointly capture rich structural relations between the tokens. For example, the depth of the LCA reflects the length of their shared prefix along the item paths, indicating semantic relatedness; $\delta_u$ and $\delta_v$ quantify how far each token is from the LCA, revealing their relative distance in the tree.  By comparing $\delta_u$ and $\delta_v$, we can infer which token lies higher in the hierarchy.  While more sophisticated strategies, \eg directly encoding the semantic embeddings of the path nodes along the tree, could carry richer information, they would incur substantial computational cost. 

With the three-dimensional characterization of token structural relations, we encode this information into a learnable bias that captures the semantic dependencies among tokens. Specifically, we revise the original self-attention scores by adding a topology-aware bias:
\begin{equation}
    \label{eq:tre}
    \tilde A_{u,v}= \text{Softmax}_v \left( \frac{(e_u W^Q)(e_v W^K)^\top}{\sqrt{d}} + \textcolor{red!80!black}{B(d^{LCA}_{u,v}, \delta_{u}, \delta_{v})} \right)
\end{equation}
where $ B(d^{LCA}_{u,v}, \delta_{u}, \delta_{v})$ denotes a learnable scalar term indexed by the triple $(d^{LCA}_{u,v}, \delta_{u}, \delta_{v})$. The revised attention score $\tilde A_{u,v}$ thus combines the original self-attention score with the topology-aware bias, effectively injecting topological information into the transformer. For efficiency, the same parameter set $B$ is shared across layers, so $B$ can be computed once and reused rather than recomputed at each layer.

\subsection{Discussions}
\label{sec:discussion}
TrieRec has the following properties:

\textbf{Model-agnostic.} Notably, our positional encodings are agnostic to the concrete tokenization model or generative model. Thus, TrieRec can be easily plugged into a variety of recent methods by adding two simple terms to the token embeddings and attention computations (as in Eq.(\ref{eq:tae}) and Eq.(\ref{eq:tre})), yielding improvements by integrating the trie information. In our experiments, we implemented TrieRec with three representative generative models, e.g., TIGER, CoST, and LETTER, and observed consistent and significant improvements.

\textbf{Efficiency.} TrieRec only incurs limited additional computational cost. Considering a user sequence with $T$ items and $T\cdot L$ tokens, and a transformer model with $d$ dimemsional embeddings, $h$ heads and $k$ layers. The TAE involves the aggregation of each tokens with their hierarchal context with MLP; the cost for generating each token encoding is $O(L^2d^2)$.  The overall cost of TAE for the sequence is $O(TL^2d^2)$, considering we do not require to recompute $B$ for tokens sharing same token path. TRE requires computing the LCA for each token pair in the sequence, with complexity $O(\log L)$ per pair, which can be further accelerated via memorizing reduplication. The computation of $B(d_{u,v}^{LCA}, \delta_{u}, \delta_{v})$ only involves index with $O(1)$.  The overall TRE complexity is $O(h T^2 L^2 \log L)$. These costs are typically substantially smaller than the baseline Transformer operations $O(k h T^2 L^2 d)$, considering $k\cdot h \cdot T$ is typically much larger than $d$, and $k \cdot d$ is also much larger than $\log L$ in practice. Our experiments show TrieRec incurs only a small computational overhead (less than 10\%) as illustrated in Figure \ref{fig:efficiency}. 

\textbf{Hyperparameter-free.} Notably, TrieRec does not introduce additional hyperparameters requiring extensive tuning. This makes TrieRec friendly for direct application.

\textbf{Differences from other positional encodings:} 1) Compared with conventional linear positional encodings \cite{attention,t5} that measure token distance along the sequence, TrieRec explicitly encodes the trie structure. Token relations are measured along the tree, enabling explicit injection of information such as LCA, tree distances, and item-path, which often yields better performance. 2) There exists another concurrent work (available only on arXiv) CoFiRec that integrates a positional encoding signaling the layer of the token. Compared with this depth-based encoding, TrieRec offers richer structural information including the surrounding context and hierarchical relations, leading to typically superior performance.

\textbf{Differences from DiscRec.} There is another concurrent work  DiscRec (available only on arXiv) that explicitly encodes information from other tokens of the same item to enrich the target token. TrieRec differs from DiscRec in several aspects: 

(i) attention in DiscRec is a linear embedding combination of tokens and does not perform finer-grained dimension-level interactions. (ii) Beyond embedding-based augmentation, our TAE also considers related structural relations to guide attention aggregation beyond tokens on a single item.

\section{Experiments}
In this section, we conduct extensive experiments to answer the following research questions: 
\begin{itemize}[leftmargin=*]
\item \textbf{RQ1:} Does \ours outperform state-of-the-art GR methods? 
\item \textbf{RQ2:} How do two positional encodings (TAE and TRE) contribute to the overall performance? 
\item \textbf{RQ3:} What is the computational overhead incurred by TrieRec?
\end{itemize}

\subsection{Experimental Setup}
\subsubsection{Datasets}
We evaluate \ours on four widely-used real-world datasets: \textbf{Amazon-Beauty}, \textbf{Amazon-Toys}, and \textbf{Amazon-Food} are three datasets that contain user ratings on products from the corresponding categories on the Amazon e-commerce platform \cite{mcauley2013amateurs}; \textbf{MovieLens-1M} is a movie rating dataset collected by the MovieLens platform \cite{harper2015movielens}.
For fair comparisons, we closely follow the setting of previous work \cite{tiger,letter}, and adopt a standard 5-core and retain only the most recent 20 interactions for each user. 
We adopt the leave-one-out strategy for data splitting, where the most recent item of each user is used for testing, the second most recent item for validation, and the remaining items for training. Detailed statistics of the datasets are summarized in Table \ref{tab:datasets}.


\begin{table}[h]
\centering
\caption{Statistics of the datasets.}
\label{tab:datasets}
\begin{tabular}{lrrrr}
\toprule
\textbf{Dataset} & \textbf{\#Users} & \textbf{\#Items} & \textbf{\#Interactions} \\ \midrule
Amazon Beauty & 22,332 & 12,086 & 198,215 \\
Amazon Toys & 19,124 & 11,758 & 165,247 \\
Amazon Food & 14,631 & 8,686 & 150,809\\
MovieLens-1M & 6,040 & 3,416 & 999,611 \\ \bottomrule
\end{tabular}
\end{table}

\begin{table*}[t]
\label{overall}
\centering
\caption{Overall performance comparison. The best results are \textbf{bolded}. Improv. indicates the relative gain of \ours over the vanilla backbone. Note that N and R denote NDCG and Recall, respectively. }
\label{tab:main_results}
\resizebox{\textwidth}{!}{
\begin{tabular}{l|cccc|cccc|cccc|cccc}
\toprule
\multirow{2}{*}{\textbf{Model}} & \multicolumn{4}{c}{\textbf{Amazon Beauty}} & \multicolumn{4}{c}{\textbf{Amazon Toys}} & \multicolumn{4}{c}{\textbf{Amazon Food}} & \multicolumn{4}{c}{\textbf{MovieLens-1M}} \\ \cmidrule(lr){2-5} \cmidrule(lr){6-9} \cmidrule(lr){10-13} \cmidrule(lr){14-17} 
 & N@5 & N@10 & R@5 & R@10 & N@5 & N@10 & R@5 & R@10 & N@5 & N@10 & R@5 & R@10 & N@5 & N@10 & R@5 & R@10 \\ \midrule
 Caser & 0.0116 & 0.0137 & 0.0163 & 0.0227 & 0.0112 & 0.0135 & 0.0161 & 0.0234 & 0.0121 & 0.0171 & 0.0189 & 0.0343 & 0.0134 & 0.0217 & 0.0217 & 0.0482 \\ 
 GRU4Rec & 0.0055 & 0.0079 & 0.0088 & 0.0163 & 0.0045 & 0.0066 & 0.0066 & 0.0131 & 0.0111 & 0.0154 & 0.0178 & 0.0312 & 0.0128 & 0.0199 & 0.0202 & 0.0417 \\ 
SASRec & 0.0177 & 0.0263 & 0.0346 & 0.0612 & 0.0229 & 0.0320 & 0.0458 & 0.0739 & 0.0210 & 0.0309 & 0.0411 & 0.0716 & 0.0440 & 0.0641 & 0.0765 & 0.1391\\
BERT4Rec & 0.0196 & 0.0269 & 0.0317 & 0.0545 & 0.0207 & 0.0265 & 0.0326 & 0.0507 & 0.0245 & 0.0334 & 0.0399 & 0.0678 & 0.0420 & 0.0597 & 0.0679 & 0.1232 \\
HSTU & 0.0176 & 0.0264 & 0.0347 & 0.0622 & 0.0238 & 0.0328 & 0.0476 & 0.0754 & 0.0231 & 0.0310 & 0.0416 & 0.0662 & 0.0511 & 0.0719 & 0.0902 & 0.1553 \\ 
CofiRec & 0.0287 & 0.0356 & 0.0430 & 0.0642 & 0.0303 & 0.0372 & 0.0449 & 0.0662 & 0.0305 & 0.0387 & 0.0467 & 0.0721 & 0.0760 & 0.0953 & 0.1129 & 0.1768 \\ 

\midrule 
TIGER & 0.0318 & 0.0395 & 0.0478 & 0.0719 & 0.0332 & 0.0418 & 0.0501 & 0.0765 & 
0.0337 & 0.0424 & 0.0519 & 0.0792 & 
0.0713 & 0.0912 & 0.1096 & 0.1713 \\
+ HiT & 0.0320 & 0.0396 & 0.0467 & 0.0705 & 0.0320 & 0.0398 & 0.0488 & 0.0732 & 0.0327 & 0.0405 & 0.0504 & 0.0746 & 0.0704 & 0.0905 & 0.1071 & 0.1697 \\
+ HiROPE & 0.0307 & 0.0385 & 0.0450 & 0.0692 & 0.0302 & 0.0383 & 0.0455 & 0.0704 & 0.0330 & 0.0431 & 0.0514 & 0.0828 & 0.0693 & 0.0896 & 0.1076 & 0.1707 \\ 
+ DiscRec & 0.0322 & 0.0395 & 0.0475 & 0.0700& 0.0351 & 0.0430 & 0.0521 & 0.0765 & 
0.0336 & 0.0429 & 0.0512 & 0.0800 & 
0.0760 & 0.0969 & 0.1174 & 0.1825 \\
+ LayerPE & 
0.0307 & 0.0383 & 0.0458 & 0.0696 & 
0.0326 & 0.0406 & 0.0487 & 0.0738 & 
0.0332 & 0.0422 & 0.0517 & 0.0794 & 
0.0763 & 0.0861 & 0.1030 & 0.1614 \\
\textbf{+ \ours} & 
\textbf{0.0335} & \textbf{0.0410} & \textbf{0.0491} & \textbf{0.0723} & 
\textbf{0.0368} & \textbf{0.0444} & \textbf{0.0530} & \textbf{0.0766 }& 
\textbf{0.0368} & \textbf{0.0460} & \textbf{0.0558} & \textbf{0.0843}& 
\textbf{0.0841} & \textbf{0.1053} & \textbf{0.1272} & \textbf{0.1929 }\\
\textit{Improv.} & \textit{+5.35\%} & \textit{+3.80\%} & \textit{+2.72\%} & \textit{+0.56\%} & \textit{+10.84\%} & \textit{+6.22\%} & \textit{+5.79\%} & \textit{+0.13\%} & \textit{+9.20\%} & \textit{+8.49\%} & \textit{+7.51\%} & \textit{+6.44\%} & \textit{+17.91\%} & \textit{+15.46\%} & \textit{+16.06\%} & \textit{+12.61\%} \\ \midrule
CoST & 0.0322 & 0.0395 & 0.0475 & 0.0703
 & 0.0331 & 0.0413 & 0.0503 & 0.0756 & 0.0335 & 0.0429 & 0.0518 & 0.0811 & 0.0719 & 0.0920 & 0.1104 & 0.1730 \\
 + HiT & 0.0312 & 0.0390 & 0.0461 & 0.0703 & 0.0312 & 0.0393 & 0.0475 & 0.0727 & 0.0327 & 0.0405 & 0.0504 & 0.0746 & 0.0723 & 0.0916 & 0.1118 & 0.1715 \\
+ HiROPE & 0.0315 & 0.0395 & 0.0468 & 0.0715 & 0.0316 & 0.0392 & 0.0486 & 0.0722 & 0.0330 & 0.0431 & 0.0514 & 0.0828 & 0.0725 & 0.0917 & 0.1119 & 0.1717 \\
+ DiscRec & 0.0316 & 0.0392 & 0.0466 & 0.0698 & 0.0332 & 0.0415 & 0.0495 & 0.0752 & 0.0344 & 0.0440 & 0.0529 & 0.0828 & 0.0785 & 0.0998 & 0.1186 & 0.1848 \\
 + LayerPE & 
0.0325 & 0.0401 & 0.0488 & \textbf{0.0722} & 
0.0326 & 0.0402 & 0.0493 & 0.0730 & 
0.0308 & 0.0400 & 0.0474 & 0.0760 & 
0.0682 & 0.0866 & 0.1048 & 0.1618 \\
\textbf{+ \ours} &
\textbf{0.0337} & \textbf{0.0409} & \textbf{0.0494} & 0.0720 & 
\textbf{0.0391} & \textbf{0.0461} & \textbf{0.0555} & \textbf{0.0771} & 
\textbf{0.0367} & \textbf{0.0461} & \textbf{0.0562} & \textbf{0.0851 }& 
\textbf{0.0847} & \textbf{0.1049} & \textbf{0.1283} & \textbf{0.1909}  \\
\textit{Improv.} & \textit{+4.66\%} & \textit{+3.54\%} & \textit{+4.00\%} & \textit{+2.42\%} & \textit{+18.13\%} & \textit{+11.62\%} & \textit{+10.34\%} & \textit{+1.98\%} & \textit{+9.55\%} & \textit{+7.46\%} & \textit{+8.49\%} & \textit{+4.93\%} & \textit{+17.80\%} & \textit{+14.02\%} & \textit{+16.21\%} & \textit{+10.35\%} \\ \midrule
LETTER & 0.0345 & 0.0427 & 0.0512 & 0.0765 & 0.0347 & 0.0430 & 0.0523 & 0.0781 & 
0.0341 & 0.0444 & 0.0526 & 0.0827 & 
0.0779 & 0.0991 & 0.1200 & 0.1858 \\
 + HiT & 0.0345 & 0.0429 & 0.0514 & 0.0774 & 0.0349 & 0.0431 & 0.0519 & 0.0777 & 0.0365 & 0.0462 & 0.0556 & 0.0849 & 0.0780 & 0.0984 & 0.1209 & 0.1843 \\
+ HiROPE & 0.0345 & 0.0428 & 0.0516 & 0.0773 & 0.0334 & 0.0421 & 0.0501 & 0.0774 & 0.0353 & 0.0450 & 0.0532 & 0.0832 & 0.0764 & 0.0967 & 0.1169 & 0.1801 \\
+ DiscRec & 
0.0352 & 0.0433 & 0.0519 & 0.0772 & 
0.0353 & 0.0437 & 0.0538 & 0.0799 & 
0.0366 & 0.0463 & 0.0557 & 0.0856 & 
0.0784 & 0.0990 & 0.1199 & 0.1841 \\
+ LayerPE & 
0.0333 & 0.0413 & 0.0495 & 0.0743 & 
0.0335 & 0.0417 & 0.0500 & 0.0754 & 
0.0329 & 0.0426 & 0.0502 & 0.0802 & 
0.0690 & 0.0905 & 0.1080 & 0.1748 \\
\textbf{+ \ours} & 
\textbf{0.0385} & \textbf{0.0464} & \textbf{0.0548} & \textbf{0.0794} & 
\textbf{0.0403} & \textbf{0.0477} & \textbf{0.0569} & \textbf{0.0801} & 
\textbf{0.0373} & \textbf{0.0469} & \textbf{0.0561} & \textbf{0.0861}& 
\textbf{0.0914} & \textbf{0.1129} & \textbf{0.1363} & \textbf{0.2031} \\
\textit{Improv.} & \textit{+11.59\%} & \textit{+8.67\%} & \textit{+7.03\%} & \textit{+3.79\%} & \textit{+16.14\%} & \textit{+10.93\%} & \textit{+8.80\%} & \textit{+2.56\%} & \textit{+9.38\%} & \textit{+5.63\%} & \textit{+6.65\%} & \textit{+4.11\%} & \textit{+17.33\%} & \textit{+13.93\%} & \textit{+13.58\%} & \textit{+9.31\%} \\ 

\bottomrule
\end{tabular}
}
\end{table*}

\subsubsection{Metrics}

We adopt two widely-used evaluation metrics in sequential recommendation: \textit{Recall@K} and \textit{NDCG@K}, with $K \in \{5, 10\}$ following existing work \cite{discrec,tiger}. For traditional recommendation methods, we predict scores of all items and select the top-K items for evaluation. For generative recommendation methods, we employ constrained beam search to ensure that the generated token sequences correspond to valid items, with the beam size set to the maximum $K$ value (\ie $10$) as in previous work \cite{wang2025msl}.




\subsubsection{Baselines}

\textbf{1) Traditional sequential recommendation models:}
These models treat items as atomic indices and focus on learning ranking functions.
\begin{itemize}[leftmargin=*]
\item \textbf{Caser}(WSDM'2018) \cite{caser}: the CNN-based method that applies convolutional filters to capture multi-level sequential patterns.
\item \textbf{GRU4Rec}(CIKM'2016) \cite{gru4rec}: the RNN-based method that leverages GRU to model the temporal dynamics of user interests.
\item \textbf{SASRec}(ICDM'2018) \cite{sasrec}: the self-attentive model that employs a causal mask Attention architecture to capture user interests.
\item \textbf{Bert4Rec:}(CIKM'2019)\cite{bert4rec} the bidirectional attention-based model that utilizes BERT to model user preference.
\end{itemize}
        
\textbf{2) Generative recommendation models:} This group includes state-of-the-art generative models including:
\begin{itemize}[leftmargin=*]
\item \textbf{HSTU}(ICML'2024) \cite{hstu}: the high-performance temporal unit that utilizes an enhanced attention mechanism.
\item \textbf{TIGER}(NeurIPS'2023) \cite{tiger}: the classic generative model that uses RQ-VAE to quantize item text embeddings into semantic IDs and performs recommendation by generating these IDs.
\item \textbf{LETTER}(CIKM'2024) \cite{letter}: the state-of-the-art method based on TIGER that incorporates collaborative signals and diversity regularization during the RQ-VAE training stage.
\item \textbf{CoST}(RecSys'2024) \cite{cost}: the state-of-the-art method based on TIGER that replaces the reconstruction loss of RQ-VAE with a contrastive loss.
\item \textbf{CofiRec}(ArXiv'2025) \cite{cofirec}: the state-of-the-art method that adopts a coarse-to-fine tokenization to generate semantic IDs, and introduces item-based encoding and a hierarchical loss.
\end{itemize}

To evaluate the backbone-robustness of \ours, we evaluate \ours with three backbones: TIGER \cite{tiger}, LETTER \cite{letter}, and CoST \cite{cost}. The selection is because they represent three distinct hierarchical quantization paradigms. Besides, they are either classical or represent SOTA methods. 

\textbf{3) Positional encoding methods:} We also compare \ours with recent positional encoding strategies on multiple backbones:
\begin{itemize}[leftmargin=*]
\item \textbf{HiT}(ICPC'2023) \cite{hit} and \textbf{HiRoPE}(ACL'2024) \cite{hirope}: Two tree-aware positional encodings proposed in other domains inject structural information through path embeddings or extended 2D RoPE. 

 \item \textbf{DiscRec}(ArXiv'2025) \cite{discrec}: a strategy that explicitly encodes information from other intra-item tokens to enrich the target token. While DiscRec involves other not positional encoding modules, we also include them for strict comparisons.
\item \textbf{LayerPE}: CofiRec \cite{cofirec} also introduces a positional encoding that signals the token layer and the item position within the sequence. We also evaluate the effect of this encoding.
\end{itemize}

\subsubsection{Implementation Details}

For all generative models, we strictly follow the configuration of TIGER \cite{tiger}. In the item tokenization stage, we use Sentence-T5-base to extract textual features and employ RQ-VAE to generate semantic IDs consisting of three semantic tokens and one collision token. In the autoregressive generation stage, we adopt a 4-layer T5 architecture as the backbone. The learning rate of the generative recommender is tuned within $\{1e^{-3}, 2e^{-3}, 3e^{-3}, 4e^{-3}, 5e^{-3}\}$ and weight decay within $\{0.1, 0.2, 0.3, 0.4\}$. All models use AdamW as the optimizer. 
More detailed hyperparameter settings are provided in the Appendix \ref{sec:appendix_hypers}.

For all traditional sequential methods, we implement and search for the optimal hyperparameters strictly following the original paper. We have traversed and frequently expanded upon the entire hyperparameter space suggested by the authors to ensure all compared methods achieve optimal performance.

\subsection{Overall Performance (RQ1)}
\label{sec:overall}
The performance comparison of our \ours and all baselines is shown in Table \ref{tab:main_results} and Table \ref{tab:rq4_comparison}. 
Our proposed \ours consistently outperforms all baselines across different metrics and datasets, demonstrating the effectiveness of explicitly encoding the trie structure in the autoregressive modeling of generative recommendation.

\textbf{Comparing with other recommendation models.}
Among all baseline methods, generative recommendation models consistently outperform traditional sequential recommendation models across all datasets. This performance gap highlights the advantages of the generative paradigm, which leverages hierarchical semantic IDs to capture both item semantics and collaborative patterns, enabling more accurate next-item prediction through structured sequence generation rather than direct ranking.
More importantly, when integrated with different generative backbones (TIGER, CoST, LETTER), \ours consistently achieves the best performance, demonstrating the effectiveness of our trie-aware positional encodings in enhancing generative recommendation.


\textbf{Comparing with other positional encoding methods.}
When integrated with different generative backbones, \ours delivers significant and stable performance improvements, with particularly notable gains on smaller $K$ values, indicating that \ours's effectiveness in identifying the most relevant items. Compared to other positional encoding methods, \ours shows superior performance for several reasons: (i) HiT and HiRoPE, originally designed for code generation tasks, lack the specific adaptations needed for generative recommendation scenarios; (ii) While LayerPE and DiscRec are tailored for generative recommendation, they rely on coarse-grained structural awareness --- LayerPE only adds local labels to input embeddings, and DiscRec achieves item-level isolation through binary masks. In contrast, our method implements fine-grained trie structure reconstruction through TAE and TRE encodings, demonstrating that explicitly reconstructing the trie's topological structure is more effective than the implicit or discrete structural marking approaches adopted in previous work.

\textbf{Comparing with trie-aware loss.} SEATER is the only trie-aware GR method that incorporates trie information through a contrastive loss. We compare our generative recommender with SEATER using the SEATER backbone \footnote{SEATER is relatively complex and difficult to integrate with other backbones.}, as shown in Table \ref{tab:rq4_comparison}. Our method achieves significantly better performance than SEATER. It indicates that simply incorporating trie information through auxiliary loss objectives is insufficient for effective trie-aware modeling.
Furthermore, SEATER focuses solely on local parent-child relationships while neglecting global positional information and higher-order structural relationships. \ours achieves a more comprehensive trie encoding strategy through TAE and TRE, yielding superior performance.

\subsection{Ablation Study(RQ2)}\label{sec:ablation}

Table \ref{tab:ablation} presents comprehensive ablation studies using LETTER as the backbone model. We systematically evaluate the contribution of each component in \ours by removing or modifying key modules.

\textbf{Effects of TAE.} Removing the entire TAE module (\textit{w/o TAE}) leads to substantial performance degradation across all metrics, demonstrating that explicitly encoding the path from root to each item node is crucial for capturing the hierarchical semantics in the trie structure. When we restrict the MLP interaction to the same feature dimension across different depths (\textit{TAE-Blocked}, where dimension $j$ at depth $k$ can only interact with dimension $j$ at other depths), performance also decreases notably, indicating that cross-dimensional interaction is also important. Most strikingly, replacing TAE with a prefix sum-based integration mechanism (\textit{TAE-Prefix}) results in the worst performance, even underperforming the original LETTER baseline, suggesting that simple cumulative integration cannot capture the complex hierarchical relationships in the trie structure, and our structured depth-aware path integration is necessary for effective trie encoding.

\textbf{Effects of TRE.} Removing the TRE module (\textit{w/o TRE}) while retaining only the standard 1D relative position bias causes significant performance drops, confirming that modeling trie-specific topological relationships is vital for generative recommendation. Further analysis reveals that both components of TRE are necessary: removing the LCA depth dimension while keeping only vertical distances (\textit{TRE-w/o LCA}) degrades performance, indicating that the absolute hierarchical level of common ancestors provides important structural context beyond simple relative distances. Similarly, using only LCA depth without relative distance components (\textit{TRE-w/o $\delta$}) also hurts performance, demonstrating that knowing the exact path lengths between nodes is equally important. These results confirm that both the common prefix information and the precise topological distances work synergistically to provide comprehensive structural awareness in the trie.

\begin{table*}[htbp]
\centering
\caption{Ablation study on four datasets (Backbone: LETTER).}
\label{tab:ablation}
\tabcolsep=3pt 
\resizebox{\textwidth}{!}{ 
\begin{tabular}{lcccccccccccccccc}
\toprule
\multirow{2}{*}{\textbf{Variant}} & \multicolumn{4}{c}{\textbf{Amazon Beauty}} & \multicolumn{4}{c}{\textbf{Amazon Toy}} & \multicolumn{4}{c}{\textbf{Amazon Food}} & \multicolumn{4}{c}{\textbf{MovieLens-1M}} \\ 
\cmidrule(lr){2-5} \cmidrule(lr){6-9} \cmidrule(lr){10-13} \cmidrule(lr){14-17}
 & N@5 & N@10 & R@5 & R@10 & N@5 & N@10 & R@5 & R@10 & N@5 & N@10 & R@5 & R@10 & N@5 & N@10 & R@5 & R@10 \\ \midrule
\textbf{\ours} & \textbf{0.0385} & \textbf{0.0464} & \textbf{0.0548} & \textbf{0.0794} & \textbf{0.0403} & \textbf{0.0477} & \textbf{0.0569} & \textbf{0.0801} & \textbf{0.0373} & \textbf{0.0469} & \textbf{0.0561} & \textbf{0.0861} & \textbf{0.0914} & \textbf{0.1129} & \textbf{0.1363} & \textbf{0.2031} \\ \midrule
(1) w/o TAE & 0.0360 & 0.0442 & 0.0528 & 0.0781 & 0.0383 & 0.0464 & 0.0562 & 0.0800 & 0.0361 & 0.0461 & 0.0539 & 0.0847 & 0.0740 & 0.0946 & 0.1138 & 0.1779 \\
(2) w/o TRE & 0.0363 & 0.0444 & 0.0527 & 0.0779 & 0.0381 & 0.0459 & 0.0549 & 0.0791 & 0.0348 & 0.0438 & 0.0524 & 0.0803 & 0.0847 & 0.1061 & 0.1287 & 0.1951  \\
(3) TAE-Blocked & 0.0362 & 0.0450 & 0.0531 & 0.0790 & 0.0375 & 0.0454 & 0.0559 & 0.0799 & 0.0346 & 0.0441 & 0.0531 & 0.0826 & 0.0762 & 0.0971 & 0.1166 & 0.1816 \\ 
(4) TAE-Prefix & 0.0343 & 0.0425 & 0.0513 & 0.0766 & 0.0339 & 0.0418 & 0.0515 & 0.0760 & 0.0351 & 0.0444 & 0.0534 & 0.0824 & 0.0734 & 0.0956 & 0.1127 & 0.1814\\
(5) TRE-w/o LCA & 0.0374 & 0.0453 & 0.0540 & 0.0787 & 0.0385 & 0.0465 & 0.0554 & 0.0799 & 0.0355 & 0.0455 & 0.0527 & 0.0837 & 0.0901 & 0.1112 & 0.1349 & 0.2007\\
(6) TRE-w/o $\delta$ & 0.0371 & 0.0449 & 0.0543 & 0.0786 & 0.0397 & 0.0473 & 0.0566 & 0.0795 & 0.0372 & 0.0468 & 0.0551 & 0.0850 & 0.0900 & 0.1118 & 0.1343 & 0.2017\\
\bottomrule
\end{tabular}
}
\end{table*}

\begin{table}[h]
\centering
\caption{Performance comparison between TrieRec and SEATER (Backbone: LETTER).}
\label{tab:rq4_comparison}
\small
\begin{tabular}{llcccc}
\toprule
\textbf{Dataset} & \textbf{Model} & \textbf{N@5} & \textbf{N@10} & \textbf{R@5} & \textbf{R@10} \\
\midrule
\multirow{2}{*}{\textbf{Beauty}} & SEATER & 0.0334 & 0.0413 & 0.0508 & 0.0572 \\
& \textbf{\ours} & \textbf{0.0385} & \textbf{0.0464} & \textbf{0.0548} & \textbf{0.0794} \\
\midrule
\multirow{2}{*}{\textbf{Toy}} & SEATER & 0.0351 & 0.0432 & 0.0523 & 0.0778 \\
& \textbf{\ours} & \textbf{0.0403} & \textbf{0.0477} & \textbf{0.0569} & \textbf{0.0801} \\
\midrule
\multirow{2}{*}{\textbf{Food}} & SEATER & 0.0342 & 0.0440 & 0.0516 & 0.0822 \\
& \textbf{\ours} & \textbf{0.0373} & \textbf{0.0469} & \textbf{0.0561} & \textbf{0.0861} \\
\midrule
\multirow{2}{*}{\textbf{ML-1M}} & SEATER & 0.0726 & 0.0948 & 0.1123 & 0.1813 \\
& \textbf{\ours} & \textbf{0.0914} & \textbf{0.1129} & \textbf{0.1363} & \textbf{0.2031} \\
\bottomrule
\end{tabular}
\end{table}
\subsection{Efficiency Analysis (RQ3)}
\begin{figure}[htbp]
    \centering
    \includegraphics[width=\linewidth]{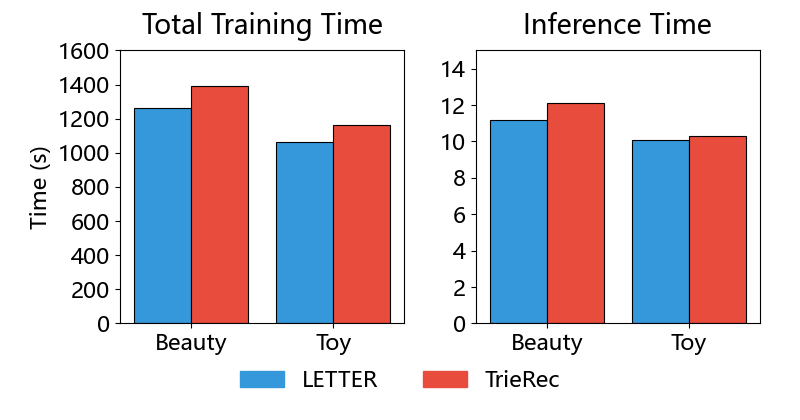}
    \caption{Efficiency comparison between \ours and baseline LETTER on Beauty and Toy datasets. The left panel shows the total training time, and the right panel shows the inference latency.}
    \label{fig:efficiency}
\end{figure}

To evaluate the practical feasibility of our proposed framework, we compare the computational efficiency of \ours against the baseline LETTER. Figure \ref{fig:efficiency} illustrates the total training time and single inference latency on two representative datasets.

The results indicate that while TrieRec introduces additional components to enhance recommendation quality, it remains within a comparable order of magnitude in terms of both training and inference costs. Although the sophisticated modeling in TrieRec leads to a slight increase in execution time, the growth is well-constrained and does not hinder the scalability of the model. Considering the significant performance improvements demonstrated in previous experiments, the marginal overhead is a justifiable trade-off for the substantial gains in recommendation accuracy.

\section{Related Work}
\textbf{Sequential Recommendation.}
Sequential recommendation aims to predict the next item a user is likely to interact with based on their historical behavior sequences. Early neural models explored various deep learning architectures to identify temporal patterns and long-term dependencies within interaction histories, such as GRU4Rec \cite{gru4rec}, Caser \cite{caser}, SASRec \cite{sasrec} and Bert4Rec \cite{bert4rec}. Recently, there emerge diverse research paths to further enhance recommendation performance, 
such as leveraging various user actions to capture fine-grained preferences \cite{wu2024multi,su2023personalized} or incorporating additional item side information \cite{zhou2023bootstrap, li2023text,zhou2025disentangled}

\textbf{Generative Recommendation.}
Generative recommendation marks a significant departure from the traditional paradigm by representing items as discrete token sequences and formulating recommendation as a sequence-to-sequence decoding task. Generally, the generative recommendation paradigm can be decomposed into two fundamental stages: item tokenization and the downstream generative recommendation model.

Recently, generative recommendation has emerged as a promising paradigm by reformulating the ranking task as a sequence-to-sequence problem. TIGER \cite{tiger} serves as a pioneer in this field, utilizing RQ-VAE \cite{rqvae} for item tokenization and a T5-based architecture \cite{t5} for autoregressive generation. Following this, research has primarily diverged into two directions. The first line focuses on item indexing strategies: for instance, LETTER \cite{letter} enhances RQ-VAE by incorporating collaborative signals, while CoST \cite{cost} employs contrastive learning to derive more discriminative identifiers. Other concurrent works have also explored advanced hierarchical indexing methods or VAE-based enhancements \cite{hid-vae, simcit, unger}. The second direction attempts to optimize the downstream generative models themselves to better handle recommendation tasks \cite{hstu, car, cobra, mbgr}. More recently, some studies have shifted toward capturing hierarchical information directly from item codes. For example, CoFiRec \cite{cofirec} introduces item-based position encoding, and DiscRec \cite{discrec} integrates both collaborative and semantic information at the item level. Despite these advancements, most existing models still treat item codes as flattened 1D sequences during the generation process. This approach inevitably loses the critical topological information in the hierarchical item tree.

\textbf{Tree-structure Modeling.} 
In the recommendation field, tree structures are widely recognized for their coarse-to-fine properties and have been extensively utilized for hierarchical indexing and retrieval. Classical approaches such as TDM \cite{tdm} and JTM \cite{jtm} organize large-scale item catalogs into tree-based indexes to enable efficient logarithmic-time retrieval. Recently, tree structures have further been exploited for layer-wise retrieval \cite{llmtreerec}, transforming the recommendation task into a path-searching process on a taxonomy. In the generative recommendation paradigm, SEATER \cite{seater} attempts to model the tree hierarchy by introducing shared-prefix-aware and parent-child relationship-aware losses into the objective function. However, such optimization-level constraints merely provide external guidance during training, failing to fully and explicitly capture the trie structure within the model's computation. As discussed in SubSection \ref{sec:discussion}, different from these works, we incorporate the trie structure into the Transformer as specialized position encodings. This design is more natively compatible with current generative tree recommendation frameworks and demonstrates superior effectiveness.

Modeling hierarchical structures is also a long-standing challenge in AI. Beyond the recommendation domain, various Tree-aware neural network have been proposed to capture the inductive bias of tree topologies. Early efforts in NLP, such as Tree-LSTM \cite{treelstm} and Tree-CNN \cite{treecnn}, attempted to process hierarchical syntax trees through recursive or convolutional structures. With the dominance of Transformers, researchers have developed specialized encodings to inject structural information into self-attention. These include methods based on path information \cite{pathtree,hit,TPtrans}, such as HiT encode the trajectory from the root, and those integrating node positions and relative topological distances \cite{hirope, treetransformer}, such as HiRoPE which utilizes 2D Rotary Positional Embedding (RoPE) to inject tree structure information. These methods were designed for code representation tasks, aiming to capture program-specific structural semantics, such as variable scoping within nested blocks or interaction patterns within function-level units. Specifically, these methods focus on syntax trees, which differ inherently from the prefix tree structure in recommendation. However, their methods are not suitable for the recommendation scenario where the entire item tokens together form a Trie. They struggle to capture key information such as the hierarchical semantics of identifiers and the structural correlations.  Consequently, we also observe their performance are compromised as shown in Table \ref{tab:main_results}.

\section{Conclusion}

This paper presented \ours, a trie-aware generative recommendation method that addresses the critical limitation of existing approaches overlooking the inherent tree structure during autoregressive modeling. By introducing Trie-aware Absolute Positional Encoding
(TAE) and Topology-aware Relative Positional
Encoding (TRE) as complementary positional encodings, TrieRec enables Transformers to leverage rich structural and topological information, achieving an average relative improvement of 8.83\% across three GR backbones on four real-world datasets with minimal computational overhead. It highlights the importance of structural inductive bias in generative recommendation and opens promising directions for future research in structure-aware generative recommendation modeling.
In the future, it would be of great interest to improve inference efficiency through structure-aware decoding strategies and extend TrieRec to multi-modal and cross-domain recommendation scenarios where hierarchical item representations naturally emerge.

\begin{appendices}
\section{Implementation Details}
\label{sec:appendix_hypers}
For the generative recommender's architecture, the model dimension is set to 128 with 6 attention heads and a hidden MLP size of 1024, applying a dropout rate of 0.1. The RQ-VAE tokenizer utilizes codebooks with 256 embeddings of dimension 32 each, and the weighting coefficient $\beta$ is set to 0.25. During the RQ-VAE training phase, we employ K-means initialization and train for 20,000 epochs with a learning rate of $1e^{-3}$; to ensure ID uniqueness, we select the epoch with the lowest collision rate before appending the fourth digit. For CofiRec, we implement its hierarchical indexing with 4 codebooks, where the fourth codebook specifically utilizes SASRec embeddings as input without applying a collision-removal strategy. 
\end{appendices}
\begin{acks}
To Robert, for the bagels and explaining CMYK and color spaces.
\end{acks}

\bibliographystyle{ACM-Reference-Format}
\bibliography{references}

@article{tiger,
  title={Recommender systems with generative retrieval},
  author={Rajput, Shashank and Mehta, Nikhil and Singh, Anima and Hulikal Keshavan, Raghunandan and Vu, Trung and Heldt, Lukasz and Hong, Lichan and Tay, Yi and Tran, Vinh and Samost, Jonah and others},
  journal={Advances in Neural Information Processing Systems},
  volume={36},
  pages={10299--10315},
  year={2023}
}

@article{gru4rec,
  title={Session-based recommendations with recurrent neural networks},
  author={Hidasi, Bal{\'a}zs and Karatzoglou, Alexandros and Baltrunas, Linas and Tikk, Domonkos},
  journal={arXiv preprint arXiv:1511.06939},
  year={2015}
}

@inproceedings{caser,
  title={Personalized top-n sequential recommendation via convolutional sequence embedding},
  author={Tang, Jiaxi and Wang, Ke},
  booktitle={Proceedings of the eleventh ACM international conference on web search and data mining},
  pages={565--573},
  year={2018}
}

@inproceedings{sasrec,
  title={Self-attentive sequential recommendation},
  author={Kang, Wang-Cheng and McAuley, Julian},
  booktitle={2018 IEEE international conference on data mining (ICDM)},
  pages={197--206},
  year={2018},
  organization={IEEE}
}

@inproceedings{bert4rec,
  title={BERT4Rec: Sequential recommendation with bidirectional encoder representations from transformer},
  author={Sun, Fei and Liu, Jun and Wu, Jian and Pei, Changhua and Lin, Xiao and Ou, Wenwu and Jiang, Peng},
  booktitle={Proceedings of the 28th ACM international conference on information and knowledge management},
  pages={1441--1450},
  year={2019}
}

@inproceedings{letter,
  title={Learnable item tokenization for generative recommendation},
  author={Wang, Wenjie and Bao, Honghui and Lin, Xinyu and Zhang, Jizhi and Li, Yongqi and Feng, Fuli and Ng, See-Kiong and Chua, Tat-Seng},
  booktitle={Proceedings of the 33rd ACM International Conference on Information and Knowledge Management},
  pages={2400--2409},
  year={2024}
}

@inproceedings{cost,
  title={Cost: Contrastive quantization based semantic tokenization for generative recommendation},
  author={Zhu, Jieming and Jin, Mengqun and Liu, Qijiong and Qiu, Zexuan and Dong, Zhenhua and Li, Xiu},
  booktitle={Proceedings of the 18th ACM Conference on Recommender Systems},
  pages={969--974},
  year={2024}
}

@article{hid-vae,
  title={HiD-VAE: Interpretable Generative Recommendation via Hierarchical and Disentangled Semantic IDs},
  author={Fang, Dengzhao and Gao, Jingtong and Zhu, Chengcheng and Li, Yu and Zhao, Xiangyu and Chang, Yi},
  journal={arXiv preprint arXiv:2508.04618},
  year={2025}
}

@article{cofirec,
  title={CoFiRec: Coarse-to-Fine Tokenization for Generative Recommendation},
  author={Wei, Tianxin and Ning, Xuying and Chen, Xuxing and Qiu, Ruizhong and Hou, Yupeng and Xie, Yan and Yang, Shuang and Hua, Zhigang and He, Jingrui},
  journal={arXiv preprint arXiv:2511.22707},
  year={2025}
}

@article{treelstm,
  title={Improved semantic representations from tree-structured long short-term memory networks},
  author={Tai, Kai Sheng and Socher, Richard and Manning, Christopher D},
  journal={arXiv preprint arXiv:1503.00075},
  year={2015}
}

@article{treecnn,
  title={Tree-CNN: a hierarchical deep convolutional neural network for incremental learning},
  author={Roy, Deboleena and Panda, Priyadarshini and Roy, Kaushik},
  journal={Neural networks},
  volume={121},
  pages={148--160},
  year={2020},
  publisher={Elsevier}
}

@article{pathtree,
  title={Novel positional encodings to enable tree-based transformers},
  author={Shiv, Vighnesh and Quirk, Chris},
  journal={Advances in neural information processing systems},
  volume={32},
  year={2019}
}

@inproceedings{hit,
  title={Implant global and local hierarchy information to sequence based code representation models},
  author={Zhang, Kechi and Li, Zhuo and Jin, Zhi and Li, Ge},
  booktitle={2023 IEEE/ACM 31st International Conference on Program Comprehension (ICPC)},
  pages={157--168},
  year={2023},
  organization={IEEE}
}

@article{TPtrans,
  title={Integrating tree path in transformer for code representation},
  author={Peng, Han and Li, Ge and Wang, Wenhan and Zhao, Yunfei and Jin, Zhi},
  journal={Advances in Neural Information Processing Systems},
  volume={34},
  pages={9343--9354},
  year={2021}
}

@inproceedings{TreeTransformer,
  title={Rethinking positional encoding in tree transformer for code representation},
  author={Peng, Han and Li, Ge and Zhao, Yunfei and Jin, Zhi},
  booktitle={Proceedings of the 2022 conference on empirical methods in natural language processing},
  pages={3204--3214},
  year={2022}
}

@article{hirope,
  title={Hirope: Length extrapolation for code models using hierarchical position},
  author={Zhang, Kechi and Li, Ge and Zhang, Huangzhao and Jin, Zhi},
  journal={arXiv preprint arXiv:2403.19115},
  year={2024}
}

@inproceedings{tdm,
  title={Learning tree-based deep model for recommender systems},
  author={Zhu, Han and Li, Xiang and Zhang, Pengye and Li, Guozheng and He, Jie and Li, Han and Gai, Kun},
  booktitle={Proceedings of the 24th ACM SIGKDD international conference on knowledge discovery \& data mining},
  pages={1079--1088},
  year={2018}
}

@article{jtm,
  title={Joint optimization of tree-based index and deep model for recommender systems},
  author={Zhu, Han and Chang, Daqing and Xu, Ziru and Zhang, Pengye and Li, Xiang and He, Jie and Li, Han and Xu, Jian and Gai, Kun},
  journal={Advances in Neural Information Processing Systems},
  volume={32},
  year={2019}
}

@inproceedings{llmtreerec,
  title={Llmtreerec: Unleashing the power of large language models for cold-start recommendations},
  author={Zhang, Wenlin and Wu, Chuhan and Li, Xiangyang and Wang, Yuhao and Dong, Kuicai and Wang, Yichao and Dai, Xinyi and Zhao, Xiangyu and Guo, Huifeng and Tang, Ruiming},
  booktitle={Proceedings of the 31st International Conference on Computational Linguistics},
  pages={886--896},
  year={2025}
}

@inproceedings{seater,
  title={Generative retrieval with semantic tree-structured identifiers and contrastive learning},
  author={Si, Zihua and Sun, Zhongxiang and Chen, Jiale and Chen, Guozhang and Zang, Xiaoxue and Zheng, Kai and Song, Yang and Zhang, Xiao and Xu, Jun and Gai, Kun},
  booktitle={Proceedings of the 2024 Annual International ACM SIGIR Conference on Research and Development in Information Retrieval in the Asia Pacific Region},
  pages={154--163},
  year={2024}
}

@article{discrec,
  title={DiscRec: Disentangled Semantic-Collaborative Modeling for Generative Recommendation},
  author={Liu, Chang and Bai, Yimeng and Zhao, Xiaoyan and Zhang, Yang and Feng, Fuli and Rong, Wenge},
  journal={arXiv preprint arXiv:2506.15576},
  year={2025}
}

@article{hstu,
  title={Actions speak louder than words: Trillion-parameter sequential transducers for generative recommendations},
  author={Zhai, Jiaqi and Liao, Lucy and Liu, Xing and Wang, Yueming and Li, Rui and Cao, Xuan and Gao, Leon and Gong, Zhaojie and Gu, Fangda and He, Michael and others},
  journal={arXiv preprint arXiv:2402.17152},
  year={2024}
}

@article{car,
  title={Act-With-Think: Chunk Auto-Regressive Modeling for Generative Recommendation},
  author={Wang, Yifan and Gan, Weinan and Xiao, Longtao and Zhu, Jieming and Chang, Heng and Wang, Haozhao and Zhang, Rui and Dong, Zhenhua and Tang, Ruiming and Li, Ruixuan},
  journal={arXiv preprint arXiv:2506.23643},
  year={2025}
}

@inproceedings{rqvae,
  title={Autoregressive image generation using residual quantization},
  author={Lee, Doyup and Kim, Chiheon and Kim, Saehoon and Cho, Minsu and Han, Wook-Shin},
  booktitle={Proceedings of the IEEE/CVF conference on computer vision and pattern recognition},
  pages={11523--11532},
  year={2022}
}

@article{simcit,
  title={A Simple Contrastive Framework Of Item Tokenization For Generative Recommendation},
  author={Zhai, Penglong and Yuan, Yifang and Di, Fanyi and Li, Jie and Liu, Yue and Li, Chen and Huang, Jie and Wang, Sicong and Xu, Yao and Li, Xin},
  journal={arXiv preprint arXiv:2506.16683},
  year={2025}
}

@article{cobra,
  title={Sparse meets dense: Unified generative recommendations with cascaded sparse-dense representations},
  author={Yang, Yuhao and Ji, Zhi and Li, Zhaopeng and Li, Yi and Mo, Zhonglin and Ding, Yue and Chen, Kai and Zhang, Zijian and Li, Jie and Li, Shuanglong and others},
  journal={arXiv preprint arXiv:2503.02453},
  year={2025}
}

@inproceedings{mbgr,
  title={Multi-behavior generative recommendation},
  author={Liu, Zihan and Hou, Yupeng and McAuley, Julian},
  booktitle={Proceedings of the 33rd ACM International Conference on Information and Knowledge Management},
  pages={1575--1585},
  year={2024}
}

@article{unger,
  title={Unger: Generative recommendation with a unified code via semantic and collaborative integration},
  author={Xiao, Longtao and Wang, Haozhao and Wang, Cheng and Ji, Linfei and Wang, Yifan and Zhu, Jieming and Dong, Zhenhua and Zhang, Rui and Li, Ruixuan},
  journal={ACM Transactions on Information Systems},
  volume={44},
  number={2},
  pages={1--31},
  year={2025},
  publisher={ACM New York, NY}
}

@inproceedings{zhou2023bootstrap,
  title={Bootstrap latent representations for multi-modal recommendation},
  author={Zhou, Xin and Zhou, Hongyu and Liu, Yong and Zeng, Zhiwei and Miao, Chunyan and Wang, Pengwei and You, Yuan and Jiang, Feijun},
  booktitle={Proceedings of the ACM web conference 2023},
  pages={845--854},
  year={2023}
}

@inproceedings{li2023text,
  title={Text is all you need: Learning language representations for sequential recommendation},
  author={Li, Jiacheng and Wang, Ming and Li, Jin and Fu, Jinmiao and Shen, Xin and Shang, Jingbo and McAuley, Julian},
  booktitle={Proceedings of the 29th ACM SIGKDD Conference on Knowledge Discovery and Data Mining},
  pages={1258--1267},
  year={2023}
}

@inproceedings{zhou2025disentangled,
  title={Disentangled Graph Debiasing for Next POI Recommendation},
  author={Zhou, Hailun and Xu, Jiajie and Zhu, Qiaoming and Liu, Chengfei},
  booktitle={Proceedings of the 48th International ACM SIGIR Conference on Research and Development in Information Retrieval},
  pages={1779--1788},
  year={2025}
}

@inproceedings{wu2024multi,
  title={When Multi-Behavior Meets Multi-Interest: Multi-Behavior Sequential Recommendation with Multi-Interest Self-Supervised Learning},
  author={Wu, Binquan and Cheng, Yu and Yuan, Haitao and Ma, Qianli},
  booktitle={2024 IEEE 40th International Conference on Data Engineering (ICDE)},
  pages={845--858},
  year={2024},
  organization={IEEE}
}

@inproceedings{su2023personalized,
  title={Personalized behavior-aware transformer for multi-behavior sequential recommendation},
  author={Su, Jiajie and Chen, Chaochao and Lin, Zibin and Li, Xi and Liu, Weiming and Zheng, Xiaolin},
  booktitle={Proceedings of the 31st ACM international conference on multimedia},
  pages={6321--6331},
  year={2023}
}

@article{t5,
  title={Exploring the limits of transfer learning with a unified text-to-text transformer},
  author={Raffel, Colin and Shazeer, Noam and Roberts, Adam and Lee, Katherine and Narang, Sharan and Matena, Michael and Zhou, Yanqi and Li, Wei and Liu, Peter J},
  journal={Journal of machine learning research},
  volume={21},
  number={140},
  pages={1--67},
  year={2020}
}

@article{gpt3,
  title={Language models are few-shot learners},
  author={Brown, Tom and Mann, Benjamin and Ryder, Nick and Subbiah, Melanie and Kaplan, Jared D and Dhariwal, Prafulla and Neelakantan, Arvind and Shyam, Pranav and Sastry, Girish and Askell, Amanda and others},
  journal={Advances in neural information processing systems},
  volume={33},
  pages={1877--1901},
  year={2020}
}

@article{attention,
  title={Attention is all you need},
  author={Vaswani, Ashish and Shazeer, Noam and Parmar, Niki and Uszkoreit, Jakob and Jones, Llion and Gomez, Aidan N and Kaiser, {\L}ukasz and Polosukhin, Illia},
  journal={Advances in neural information processing systems},
  volume={30},
  year={2017}
}

@article{gpt4,
  title={Gpt-4 technical report},
  author={Achiam, Josh and Adler, Steven and Agarwal, Sandhini and Ahmad, Lama and Akkaya, Ilge and Aleman, Florencia Leoni and Almeida, Diogo and Altenschmidt, Janko and Altman, Sam and Anadkat, Shyamal and others},
  journal={arXiv preprint arXiv:2303.08774},
  year={2023}
}

@article{zhao2024recommender,
  title={Recommender systems in the era of large language models (llms)},
  author={Zhao, Zihuai and Fan, Wenqi and Li, Jiatong and Liu, Yunqing and Mei, Xiaowei and Wang, Yiqi and Wen, Zhen and Wang, Fei and Zhao, Xiangyu and Tang, Jiliang and others},
  journal={IEEE Transactions on Knowledge and Data Engineering},
  volume={36},
  number={11},
  pages={6889--6907},
  year={2024},
  publisher={IEEE}
}

@inproceedings{deldjoo2024review,
  title={A review of modern recommender systems using generative models (gen-recsys)},
  author={Deldjoo, Yashar and He, Zhankui and McAuley, Julian and Korikov, Anton and Sanner, Scott and Ramisa, Arnau and Vidal, Ren{\'e} and Sathiamoorthy, Maheswaran and Kasirzadeh, Atoosa and Milano, Silvia},
  booktitle={Proceedings of the 30th ACM SIGKDD conference on Knowledge Discovery and Data Mining},
  pages={6448--6458},
  year={2024}
}

@article{koren2009matrix,
  title={Matrix factorization techniques for recommender systems},
  author={Koren, Yehuda and Bell, Robert and Volinsky, Chris},
  journal={Computer},
  volume={42},
  number={8},
  pages={30--37},
  year={2009},
  publisher={IEEE}
}

@inproceedings{he2017neural,
  title={Neural collaborative filtering},
  author={He, Xiangnan and Liao, Lizi and Zhang, Hanwang and Nie, Liqiang and Hu, Xia and Chua, Tat-Seng},
  booktitle={Proceedings of the 26th international conference on world wide web},
  pages={173--182},
  year={2017}
}

@article{li2025survey,
  title={A Survey of Generative Recommendation from a Tri-Decoupled Perspective: Tokenization, Architecture, and Optimization},
  author={Li, Xiaopeng and Chen, Bo and She, Junda and Cao, Shiteng and Wang, You and Jia, Qinlin and He, Haiying and Zhou, Zheli and Liu, Zhao and Liu, Ji and others},
  year={2025},
  publisher={Preprints}
}

@article{wang2023generative,
  title={Generative recommendation: Towards next-generation recommender paradigm},
  author={Wang, Wenjie and Lin, Xinyu and Feng, Fuli and He, Xiangnan and Chua, Tat-Seng},
  journal={arXiv preprint arXiv:2304.03516},
  year={2023}
}

@inproceedings{luo2025qarm,
  title={Qarm: Quantitative alignment multi-modal recommendation at kuaishou},
  author={Luo, Xinchen and Cao, Jiangxia and Sun, Tianyu and Yu, Jinkai and Huang, Rui and Yuan, Wei and Lin, Hezheng and Zheng, Yichen and Wang, Shiyao and Hu, Qigen and others},
  booktitle={Proceedings of the 34th ACM International Conference on Information and Knowledge Management},
  pages={5915--5922},
  year={2025}
}

@article{vijayakumar2016diverse,
  title={Diverse beam search: Decoding diverse solutions from neural sequence models},
  author={Vijayakumar, Ashwin K and Cogswell, Michael and Selvaraju, Ramprasath R and Sun, Qing and Lee, Stefan and Crandall, David and Batra, Dhruv},
  journal={arXiv preprint arXiv:1610.02424},
  year={2016}
}

@inproceedings{zhuo2020learning,
  title={Learning optimal tree models under beam search},
  author={Zhuo, Jingwei and Xu, Ziru and Dai, Wei and Zhu, Han and Li, Han and Xu, Jian and Gai, Kun},
  booktitle={International Conference on Machine Learning},
  pages={11650--11659},
  year={2020},
  organization={PMLR}
}

@inproceedings{bao2024decoding,
  title={Decoding matters: Addressing amplification bias and homogeneity issue in recommendations for large language models},
  author={Bao, Keqin and Zhang, Jizhi and Zhang, Yang and Huo, Xinyue and Chen, Chong and Feng, Fuli},
  booktitle={Proceedings of the 2024 Conference on Empirical Methods in Natural Language Processing},
  pages={10540--10552},
  year={2024}
}

@inproceedings{wang2025msl,
  title={Msl: Not all tokens are what you need for tuning llm as a recommender},
  author={Wang, Bohao and Liu, Feng and Chen, Jiawei and Lou, Xingyu and Zhang, Changwang and Wang, Jun and Sun, Yuegang and Feng, Yan and Chen, Chun and Wang, Can},
  booktitle={Proceedings of the 48th International ACM SIGIR Conference on Research and Development in Information Retrieval},
  pages={1912--1922},
  year={2025}
}

@inproceedings{covington2016deep,
  title={Deep neural networks for youtube recommendations},
  author={Covington, Paul and Adams, Jay and Sargin, Emre},
  booktitle={Proceedings of the 10th ACM conference on recommender systems},
  pages={191--198},
  year={2016}
}

@inproceedings{bender2000lca,
  title={The LCA problem revisited},
  author={Bender, Michael A and Farach-Colton, Martin},
  booktitle={Latin American Symposium on Theoretical Informatics},
  pages={88--94},
  year={2000},
  organization={Springer}
}

@inproceedings{mcauley2013amateurs,
  title={From amateurs to connoisseurs: modeling the evolution of user expertise through online reviews},
  author={McAuley, Julian John and Leskovec, Jure},
  booktitle={Proceedings of the 22nd international conference on World Wide Web},
  pages={897--908},
  year={2013}
}

@article{harper2015movielens,
  title={The movielens datasets: History and context},
  author={Harper, F Maxwell and Konstan, Joseph A},
  journal={Acm transactions on interactive intelligent systems (tiis)},
  volume={5},
  number={4},
  pages={1--19},
  year={2015},
  publisher={Acm New York, NY, USA}
}










\end{document}